\documentclass[aps,pra,twocolumn,amsmath,amssymb,nofootinbib,showpacs,superscriptaddress, longbibliography]{revtex4-2}
\usepackage[english]{babel}
\usepackage{latexsym}
\usepackage{graphics}
\usepackage{graphicx}
\usepackage{epsfig}
\usepackage{color}
\usepackage{bm}
\usepackage{amsmath}
\usepackage{amssymb}
\usepackage{amsthm}
\usepackage{dcolumn}
\usepackage{dsfont}
\usepackage{bbm}
\usepackage{float}
\usepackage[bookmarks=true,
   colorlinks=true,
   linkcolor=blue,
   urlcolor=blue,
   citecolor=blue,
   bookmarks=true,
   hyperindex=true
]{hyperref}

\usepackage{color}
\usepackage{epstopdf}
\usepackage{cleveref}
\usepackage[svgnames]{xcolor}
\usepackage{physics}
\usepackage[sort&compress]{natbib}
\usepackage{bbold}
\usepackage{enumerate}
\usepackage{bm}% bold math

\newcommand{\be}{\begin{equation}}
\newcommand{\ee}{\end{equation}}

\newcommand{\beq}{\begin{eqnarray}}
\newcommand{\eeq}{\end{eqnarray}}

\def\H1{\widehat{H}_1}

\begin{document}

\title{One generalization of the Dicke-type models}

\author{Denis V. Kurlov}
% \affiliation{Philosophisch-Naturwissenschaftliche Fakult"{a}t
% Departement Physik, Universit\"{a}t Basel, Klingelbergstrasse 82
% 4056 Basel
% Schweiz}
\affiliation{Department of Physics, University of Basel, Klingelbergstrasse 81, CH-4056 Basel, Switzerland}
\affiliation{Russian Quantum Center, Skolkovo, Moscow 121205, Russia}
\affiliation{National University of Science and Technology ``MISIS”,  Moscow 119049, Russia}

\author{Aleksey K. Fedorov}
\affiliation{Russian Quantum Center, Skolkovo, Moscow 121205, Russia}
\affiliation{National University of Science and Technology ``MISIS”,  Moscow 119049, Russia}

\author{Alexandr Garkun}
\affiliation{The Institute of Applied Physics of the National Academy of Sciences of Belarus, Minsk 220072, Akademicheskaya 16, Belarus}
\affiliation{Department of Theoretical Physics and Astrophysics, Belarusian State University, Minsk 220030, Nezavisimosti 4, Belarus}

\author{Vladimir Gritsev}
\affiliation{Institute for Theoretical Physics, Universiteit van Amsterdam, Science Park 904, Amsterdam, The Netherlands}
\affiliation{Russian Quantum Center, Skolkovo, Moscow 121205, Russia}

\begin{abstract}
We discuss one family of possible generalizations of the Jaynes-Cummings and the Tavis-Cummings models using the technique of algebraic Bethe ansatz related to the Gaudin-type models.
In particular, we present a family of (generically) non-Hermitian Hamiltonians that generalize paradigmatic quantum-optical models.
Further directions of our research include studying physical properties of the obtained generalized models.
\end{abstract}

\maketitle
\section{Introduction}
The Jaynes-Cummings model~\cite{J-C} plays the central role in the field of quantum optics, see, e.g.,  Refs.~\cite{kilin1990quantum, scully_zubairy_1997, vogel2006quantum, gerry2005introductory} for a review. It describes an interaction between a single bosonic mode (e.g., a cavity photon of a fixed frequency $\omega$) and a single two-level system (e.g., a qubit) with an energy gap $\Delta$ as follows:  
\beq \label{H_JC}
H_{JC} =\omega a^{\dag}a+\Delta \sigma^{z}+g(a\sigma^{+}+a^{\dag}\sigma^{-}).
\label{J-C}
\eeq
Here $a^{\dag}$ and $a$ are bosonic creation and annihilation operators, respectively, and the Pauli sigma matrices $\sigma^{\pm,z}$ encode the two-level qubit degree of freedom. The photon-qubit coupling strength $g$ is assumed to be small. Note that in deriving the Hamiltonian~(\ref{H_JC}) we neglect the so-called counter-rotating terms $a^{\dag}\sigma^{+}$ and $a\sigma^{-}$, which are indeed present at the microscopic level.

A multi-qubit generalization of the Jaynes-Cummings model~ (\ref{J-C}) is known as the Tavis-Cummings model~\cite{T-C}.
In its turn, the Tavis-Cummings model can be considered as a limiting single-mode version of the celebrated multi-mode Dicke model~\cite{Dicke1954}. Notable results on its studies with the use of the Bethe ansatz have been obtained by Rupasov and Yudson in their work~\cite{Yudson1984}. This approach has been further developed by Yudson in his papers  ~\cite{Yudson1988,Yudson1985} introducing the contour representation, which is now used for a large class or quantum many-body models (see, e.g., Refs.~\cite{Andrei2014,Andrei2016}).
The Hamiltonian of the Tavis-Cummings model reads
\beq
H_{TC} = \omega a^{\dag}a+2\Delta S^{z}+g(a^{\dag}S^{+}+a^{\dag}S^{-}),
\label{T-C}
\eeq
where $S^{\pm,z}=\sum_{i=1}^{N}\sigma^{\pm,z}_{i}$ is a collective spin describing an ensemble of two-level systems placed in a region of space much smaller that the wavelength $\sim \omega^{-1}$ of bosonic mode. An exact stationary solution for the density matrix of the model~(\ref{T-C}) coupled to dissipative bath is shown to undergo a {\it nonequilibrium} first-order phase transition when the parameters of the frequency detuning or of the exciting-radiation power are changed, see Refs.~\cite{Carmichael1976, Kilin1980, Walls1980, Carmichael1980,  Puri1980, Hassan1980, Drummond1980, Kilin1982, Lawande1981, Hassan1982, Puri1983} for more details.
We also note a recent interest in the Jaynes-Cummings and Tavis-Cummings models in the context of quantum technologies~\cite{Brennecke2013, Deutsch2017,Nori2019,Sedov2020, Ricco2022,Luchnikov2022}. In particular, these and related models have been  simulated using superconducting qubtis~\cite{Wallraff2008,Gross2008,Ustinov2017}.

Although both the Jaynes-Cummings and the Tavis-Cummings models have been well-studied in various regimes, their generalizations may have non-trivial properties, which can be investigated with the use of a powerful method of the {\it algebraic Bethe ansatz}~\cite{Jurco, Ortz2004ExactlysolvableMD, Babelon_2007}, see Ref.~\cite{Slavnov2020} for a recent overview.  
Motivated by this idea, in this work we present a family of possible generalizations of the aforementioned models, which may lead to non-Hermitian Hamiltonians. We consider their basic properties and discuss certain directions for further studies. 

\section{Integrability of Gaudin-type models}

Models like (\ref{T-C}) belong to the class of the so-called Gaudin-type integrable models. These models were recently generalized in various ways by T. Skrypnyk in a series of papers, see Ref.~\cite{Skrypnyk_2022} and references therein. The central object behind integrability in these models is the existence of the so-called classical $r$-matrix that satisfies a classical version of the celebrated Yang-Baxter equation~\cite{Yang1967,Baxter1978}
\beq
[r_{12},r_{13}]+[r_{12},r_{23}]+[r_{13},r_{23}]=0
\label{cYBE},
\eeq
where the matrix $r_{ij}$ is defined on the tensor product of three Hilbert spaces $V_1 \otimes V_2 \otimes V_3$ and acts nontrivially on the subspace $V_i \otimes V_j$. Explicitly, one has $r_{12} = r \otimes 1$, $r_{23} = 1 \otimes r$, and $r_{13} = (1 \otimes P) (r \otimes 1) \otimes ( 1 \otimes P)$, where the $r$-matrix acting on $V \otimes V$ and $P$ is the permutation operator acting on two Hilbert spaces as $P(V_a \otimes V_b) = V_b \otimes V_a$.

The key identity responsible for the integrability of a model reads as follows:
\begin{multline}\label{L-L-comm}
[L(\lambda)\otimes 1,1\otimes L(\mu)]\\
 + [r(\lambda-\mu), L(\lambda)\otimes 1+1\otimes L(\mu)]=0,
\end{multline}
where the Lax matrix {\it in our context} is defined as a matrix with the operator-valued entries
\beq
L(\lambda)=\begin{pmatrix}
A(\lambda) & B(\lambda) \\
C(\lambda) & -A(\lambda)
\end{pmatrix}.
\label{L-oper}
\eeq
Here $A(\lambda)$, $B(\lambda)$, and $C(\lambda)$ are operators acting on the tensor product of bosonic and qubit degrees of freedom. The size of the $L$-matrix ($2\times 2$ in our case) defines the size of what is called the auxiliary space, which could be of arbitrary size. The relation (\ref{L-L-comm}) is an analogue of the Poisson brackets. Their associativity is guarantied by the classical Yang-Baxter equation (\ref{cYBE}). 

% Substituting the ansatz (\ref{L-oper}) into (\ref{L-L-comm}) with an 

Let us now take the $r$-matrix of the rational form $r_{ab}(\lambda)=P_{ab}/\lambda$, where $P_{ab}$ is a permutation operator acting between spaces $a$ and $b$. In the matrix form $P_{ab}$ is represented by the following $4\times 4$ matrix:
\beq
P=\begin{pmatrix}
1 & 0 & 0& 0\\
0 & 0 & 1 & 0\\
0 & 1 & 0 & 0\\
0 & 0 & 0 & 1
\end{pmatrix}.
\eeq
Then, substituting the $L$-matrix~(\ref{L-oper}) into (\ref{L-L-comm}) and using the rational $r$-matrix specified above,
% we come up with the $sl(2)$ loop algebra of the operators $A,B,C$
we find that the operators $A$, $B$, and $C$ satisfy the commutation relations of the $sl(2)$ loop algebra
\beq \label{G-alg_1}
~[B(\lambda),C(\mu)]&=&2 \frac{A(\lambda)-A(\mu)}{\lambda-\mu},\\
~[A(\lambda), B(\mu)]&=&\frac{B(\lambda)-B(\mu)}{\lambda-\mu},\\
~[A(\lambda), C(\mu)]&=&-\frac{C(\lambda)-C(\mu)}{\lambda-\mu},
\eeq
and 
\beq \label{G-alg_2}
[A(\lambda),A(\mu)]=[B(\lambda),B(\mu)]=[C(\lambda),C(\mu)]=0. 
\eeq
The following quantity
\beq
\frac{1}{2}{\rm Tr}(L^{2}(\lambda))=A^{2}(\lambda)+\frac{1}{2}(B(\lambda)C(\lambda)+C(\lambda)B(\lambda))
\label{Tr}
\eeq
has a key commutativity property: 
\beq
[{\rm Tr}(L^{2}(\lambda)), {\rm Tr}(L^{2}(\mu))]=0. 
\eeq
This quantity is a {\it generating } function for commuting Hamiltonians. 

At this point we have to specify a {\it representation} of the algebra in Eqs.~(\ref{G-alg_1}) -- (\ref{G-alg_2}). The Tavis-Cummings model (\ref{T-C}) is obtained if the following identification is made
\beq
A(\lambda)&=&\frac{2\lambda}{g^{2}}-\frac{\omega}{g^{2}}+\frac{S^{z}}{\lambda-\Delta},\nonumber\\
B(\lambda)&=&\frac{2a}{g}+\frac{S^{-}}{\lambda-\Delta},\nonumber\\
C(\lambda)&=&\frac{2a^{\dag}}{g}+\frac{S^{+}}{\lambda-\Delta}.
\label{TC-alg-rep}
\eeq
In this case~\cite{Babelon_2007} we arrive to the following expression: 
\begin{eqnarray}
    \frac{1}{2}{\rm Tr}(L^{2}(\lambda))&&=\frac{1}{g^{4}}(2\lambda-\omega)^{2}+\frac{4}{g^{2}}H_{N}+ \\
    &&\frac{2}{g^{2}(\lambda-\Delta)}H_{TC}+{\rm const},
\end{eqnarray}
where $H_{N}=a^{\dag}a+S^{z}$ is a total number of excitations that commutes with $H_{TC}$. Let us now proceed with looking for more general representations.

\section{Generalization}
One possible way of generalizing the above results is to search for a new representation of the Gaudin algebra (\ref{G-alg_1})-(\ref{G-alg_2}). We found a  faithful representation of the following form:
\begin{widetext}
\begin{eqnarray}\label{G-alg-new}
  A(\lambda)=\left[\frac{1}{2}(\alpha_{1}\alpha_{2}-\beta_{1}\beta_{2})\lambda+\rho\right]I+\frac{1}{\lambda}S^{z}, \qquad
B(\lambda)=\alpha_{1} a+\beta_{1} a^{\dag} + \frac{\gamma}{\lambda} S^{-}, \qquad
C(\lambda)=\beta_{2}a +\alpha_{2} a^{\dag}+\frac{1}{\gamma \lambda} S^{+},  
\end{eqnarray}
with no restrictions on the complex-valued parameters $\alpha_{1,2},\beta_{1,2},\rho,\gamma$.  
%Here we assume that all parameters are complex numbers.
In the above representation $I$ is the identity operator. Then, for the generating function we obtain
%\beq
%\frac{1}{2}Tr(L^{2}(\lambda))=\frac{1}{2}(\gamma\lambda+\rho)^{2}+\gamma S^{z} +\frac{\rho S^{z}}{\lambda}
%\eeq
%------------------
% \beq
% 	\frac{1}{2}{\rm Tr}(L^{2}(\lambda)) &=& \frac{1}{2} \left(\alpha _1 \alpha _2+\beta _1 \beta _2+2 \rho^2\right)
% 		+\left(\alpha _1   \alpha _2+\beta _1 \beta _2\right) a^{\dagger } a
% 		+ \alpha _2 \beta _1 \left(a^{\dagger }\right)^2 +\alpha _1 \beta _2  a^2 +  \left(\alpha _1 \alpha _2-\beta _1 \beta _2\right) S^{z}
% 		\label{zero} \\
% 		&+&\frac{\alpha _2 \gamma  a^{\dagger }S^{-}
% 			+\frac{\beta _1}{\gamma}  a^{\dagger   } S^{+}
% 			+\frac{\alpha _1 }{\gamma}a S^{+}
% 			+\beta _2 \gamma a S^{-}+2 \rho   S^z }{\lambda}
% 		\label{-1}\\
% 		&+&\frac{C_{2}}{\lambda^2} +\frac{1}{4} \lambda^2 \left(\alpha _1 \alpha _2-\beta _1 \beta _2\right)^2
% 		+\rho \lambda \left(\alpha _1 \alpha _2-\beta _1 \beta _2\right)\label{other},
%   \label{Tr-new}
%   \eeq
% where $C_{2}=(S^z)^2+\frac{1}{2}(S^{+} S^{-}+S^{-}S^{+})$ is the second Casimir invariant of $SU(2)$. 
% Note that in this Laurent-type expansion in terms of complex parameters $\lambda$ all (operator-valued) coefficients commute. As a consequence, the physical Hamiltonian can be constructed as an arbitrary linear combination of the coefficients in front of $\lambda^{k}$, for $k=0,-1,-2$. Namely, $H_{\text{phys}}=$Eq.(\ref{zero})+$\kappa$ numerator of Eq.(\ref{-1}) with some $\kappa$. Note also that the coefficients in front of $\lambda, \lambda^{2}$ in Eq.(\ref{other}) are proportional to the identity operator, the same holds for the first term in Eq.(\ref{zero}).
\begin{equation} \label{Tr-new}
    \frac{1}{2}{\rm Tr}(L^{2}(\lambda)) = H_0 + \frac{1}{\lambda} H_1 + \frac{1}{\lambda^2} C_2 + \frac{1}{4} \lambda^2 \left(\alpha _1 \alpha _2-\beta _1 \beta _2\right)^2
		+\rho \lambda \left(\alpha _1 \alpha _2-\beta _1 \beta _2\right),
\end{equation}
where 
\begin{equation}
C_{2}=(S^z)^2+\frac{1}{2}(S^{+} S^{-}+S^{-}S^{+})
\end{equation}
is the second Casimir invariant of $SU(2)$ and we denoted
\begin{align}
    H_0 &= \frac{1}{2} \left(\alpha _1 \alpha _2+\beta _1 \beta _2+2 \rho^2\right)
		+\left(\alpha _1   \alpha _2+\beta _1 \beta _2\right) a^{\dagger } a
		+ \alpha _2 \beta _1 \left(a^{\dagger }\right)^2 + \alpha_1 \beta_2  \, a^2 +  \left(\alpha _1 \alpha _2-\beta _1 \beta _2\right) S^{z} ,  \label{zero}\\
  H_1 &= \alpha _2 \gamma  a^{\dagger }S^{-}
			+\frac{\beta _1}{\gamma} \,  a^{\dagger } S^{+}
			+\frac{\alpha _1 }{\gamma} \, a \,S^{+}
			+\beta _2 \gamma \, a \,S^{-} + 2 \rho  \, S^z .
\end{align}

We note that in Eq.~(\ref{Tr-new}), which is the Laurent-type expansion in terms of complex parameters $\lambda$, all (operator-valued) coefficients commute. As a consequence, the physical Hamiltonian can be constructed as an arbitrary linear combination of the coefficients in front of $\lambda^{k}$, for $k=0,-1,-2$. Namely, $H_{\text{phys}} = H_0 + \kappa H_1$, with some $\kappa$. Also note that the coefficients in front of $\lambda, \lambda^{2}$ in Eq.(\ref{Tr-new}) are proportional to the identity operator, and the same holds for the first term in Eq.(\ref{zero}).
\end{widetext}

\section{Conclusions}
In this short note, we have constructed a family of (generically) non-Hermitian Hamiltonians generalizing famous quantum optical problems. These Hamiltonians contain terms quadratic in the bosonic operators and counter-rotating terms in the boson-qubit coupling. In the future we plan to investigate their physical properties in more detail. 

Using the known Bethe ansatz techniques, namely the separation of variables (see Refs.~\cite{Babelon_2007, Faribault2011, ElAraby2012}), we plan to further investigate the physical properties of the novel family of models. Of particular interest is the possible existence of phase transitions. One more interesting direction could be related to a possible $\mathfrak{gl}(2,\mathbb{C})$ structure hidden in this model. This observation comes from the obvious $\mathfrak{gl}(2)$ type structure in the bosonic sector of Eq. (\ref{G-alg-new}). Indeed, it can be made transparent by defining a transformation
\beq
\begin{pmatrix}
\tilde{a} \\
\tilde{a}^{\dag}
\end{pmatrix}=\begin{pmatrix}
\alpha_{1} & \beta_{1} \\
\beta_{2} & \alpha_{2}
\end{pmatrix}\begin{pmatrix}
a \\
a^{\dag}
\end{pmatrix},
\eeq
with the determinant $\alpha_{1}\alpha_{2}-\beta_{1}\beta_{2}$, which systematically appears in Eq.~(\ref{Tr-new}). It could also be interesting to study a dissipative version of the above model following the lines of Refs. \cite{Kilin1980, Kilin1982}. One more possible way of generalizing our findings is by including inhomogeneities, thus extending the results for the traditional Dicke model \cite{Babelon_2007}. In this case, in the  representation (\ref{TC-alg-rep}) one should make the following replacement
\beq
\frac{S^{\pm,z}}{\lambda-\Delta}\longrightarrow\sum_{j=1}^{N}\frac{S^{\pm,z}}{\lambda-\epsilon_{j}}
\eeq
where $N$ is the same as in (\ref{T-C}) and $\{\epsilon_{j}\}_{j=1}^{N}$ is an arbitrary set of complex numbers. The same replacement can also be made in Eq.~(\ref{G-alg-new}).

\textit{Acknowledgements.}
%\section*{Acknowledgements}
We would like to express our gratitude to  Professors Sergei Yakovlevich Kilin and Vladimir Isaakovich Yudson for valuable comments and interest in this communication.  

We thank the support by the Russian Science Foundation Grant No. 19-71-10092. The work by V.G. is part of the DeltaITP consortium, a program of the Netherlands Organization for Scientific Research (NWO) funded by the Dutch Ministry of Education, Culture and Science (OCW).

\bibliography{lib}
\end{document}